\documentclass[sigconf]{acmart}
\AtBeginDocument{%
  }

\setcopyright{acmlicensed}
\copyrightyear{2025}
\acmYear{2025}
\acmConference[Conference CHI '25]{Conference on Human Factors in Computing Systems}{April 27 2025}{Yokohama, Japan}
\begin{document}

\title{Understanding Physical Therapy Challenges for Older Adults through Mixed Reality}

\author{Jade Kandel}
\email{kandelj@cs.unc.edu}
\orcid{1234-5678-9012}
\affiliation{%
  \institution{University of North Carolina at Chapel Hill}
  \city{Chapel Hill}
  \state{North Carolina}
  \country{USA}
}

\author{Sriya Kasumarthi}
\affiliation{%
  \institution{University of North Carolina at Chapel Hill}
  \city{Chapel Hill}
  \state{North Carolina}
  \country{USA}
}
\email{kasri@unc.edu}

\author{Danielle Albers Szafir}
\affiliation{%
  \institution{University of North Carolina at Chapel Hill}
  \city{Chapel Hill}
  \state{North Carolina}
  \country{USA}
}
\email{danielle.szafir@cs.unc.edu}


\begin{abstract}
Physical therapy (PT) is crucial in helping older adults manage chronic conditions and weakening muscles, but older adults face increasing challenges that can impact their PT experience, including increased fatigue, memory loss, and mobility and travel constraints. While current technology attempts to facilitate remote care, they have limitations and are used in-practice infrequently. Mixed reality (MR) technology shows promise for addressing these challenges by creating immersive, context-aware environments remotely that previously could only be achieved in clinical settings. To bridge the gap between MR's potential and its practical application in geriatric PT, we conducted in-depth interviews with three PT clinicians and six older adult patients to  understand challenges with PT care and adherence that MR may address. Our findings inform design considerations for supporting older adults' needs through MR and outline technical requirements for practical implementation.

\end{abstract}


%\ccsdesc[500]{Do Not Use This Code~Generate the Correct Terms for Your Paper}
%\ccsdesc[300]{Do Not Use This Code~Generate the Correct Terms for Your Paper}
%\ccsdesc{Do Not Use This Code~Generate the Correct Terms for Your Paper}
%\ccsdesc[100]{Do Not Use This Code~Generate the Correct Terms for Your Paper}

\begin{teaserfigure}
  \includegraphics[width=\textwidth]{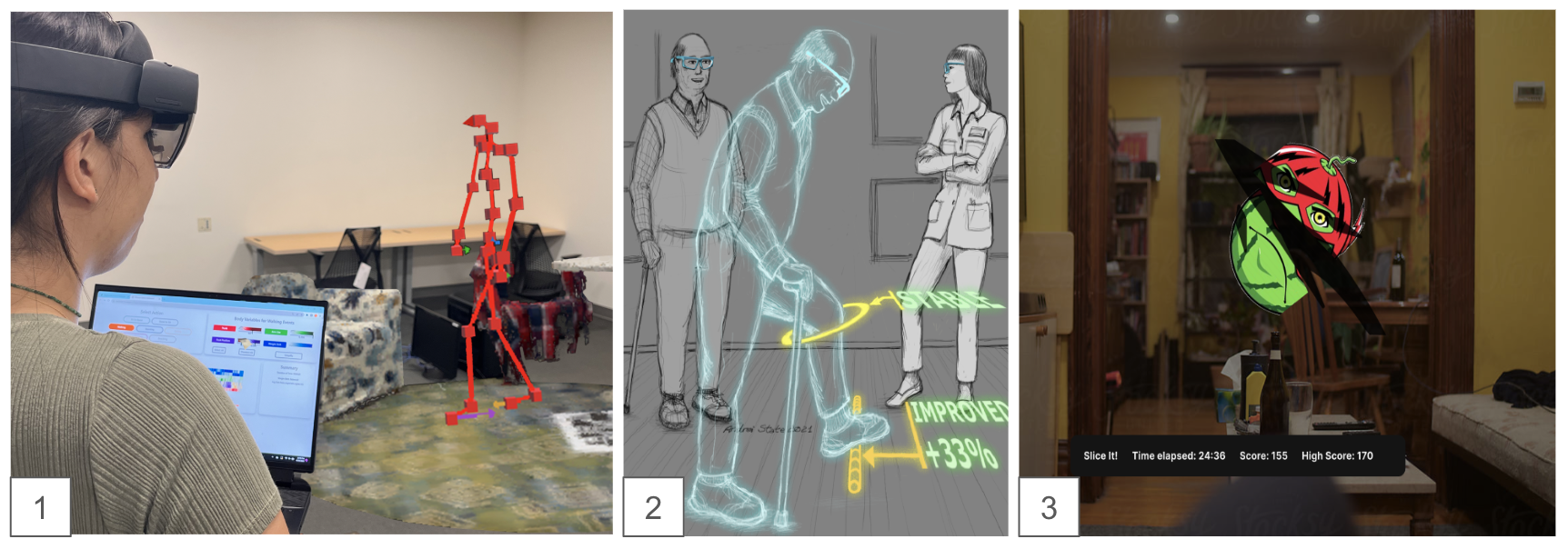}
  \caption{Examples of mixed reality assisting in multiple aspects of physical therapy patient care for older adults, including: 1) a clinician analyzing a reconstruction of the patient's body motions \cite{kandel_pd-insighter_2024}, 2) a patient and clinician discussing the patient's motor patterns, and 3) an interactive game for patients to enjoy physical therapy and increase adherence.}
  \label{fig:teaser}
\end{teaserfigure}


\maketitle
\section{Introduction}
Older adults face physiological changes including decreased muscle mass and strength, reduced bone density, impaired balance, and declining cognitive function that can impact daily activities and quality of life \cite{moustafa_motor_2016, gimigliano_world_2017, lopez-otin_hallmarks_2013}. Physical therapy (PT) can help older adults maintain mobility, mitigate muscle loss, recover from injuries and stroke, and manage chronic conditions \cite{grabli_normal_2012, shahid_comprehensive_2023, ruegsegger_health_2018}. Typical PT treatment involves weekly in-person clinic visits with guided stretches and supervised exercises \cite{merolli_use_2022}. Between visits, clinicians prescribe patients with exercises to perform at home for combating aging symptoms. However, many older adults patients struggle to maintain consistent engagement with their treatment plan despite the importance of adhering to the exercise program. Additionally, the personalized nature of physical therapy protocols--- designed to address individual's unique motor challenges---is hindered by clinicians' limited visibility into patients' physical struggles outside the clinic setting and reliance on patient's imperfect recall \cite{stone_patient_2002}.

Despite technological advances in healthcare, current PT support systems remain inadequate \cite{merolli_use_2022}. %There are currently no technological implementations used for monitoring patient's movements outside of the clinic. 
Virtual clinical visits may enable older patients to receive care from their home, and wearable sensors can provide biometrics for measuring health and activity. However, limited 2D perspective of the video call and lack of complex body motion data from sensors hinders clinician body-motion analysis and decision making. While patients can turn to video platforms for PT guidance at home, these tools fail to provide the engagement, feedback, and human connection found in in-person sessions \cite{merolli_use_2022}. %This technological gap is particularly significant for older adults, who may face unique cognitive and physical challenges that affect their ability to effectively utilize existing tools.

Mixed reality (MR) technology offers promising solutions to these challenges by enhancing both clinical understanding and patient engagement. Visual observation is fundamental to movement and posture analysis, and through immersive three-dimensional patient reconstructions, MR interfaces excel at making complex physical data intuitive for both clinicians and patients. For patients, MR provides distinct advantages over traditional desktop videos or static images by enhancing spatial perception for motion guidance. These improvements could address common challenges in memory and exercise adherence \cite{mostajeran_augmented_2020, timmermans_walking_2016, shao_acceptance_2020}.

While MR shows important promise, effectively leveraging MR for geriatric PT care requires a better understanding of older patients and clinician needs. We conducted interviews with three PT clinicians specializing in geriatric care and six older adult PT patients. From synthesizing the feedback into key challenges, we discuss how MR can serve both patients and clinicians across care environments and goals. In presenting these MR opportunities, we discuss potential design choices that can best support user needs, and highlight necessary future developments for effective real-world deployment. %given older adult described challenges and needs.
%Through understanding the challenges and needs for geriatric PT care, we present promising design considerations and opportunities for MR to improve PT care and quality of living for older adults. 

%By examining both current challenges in geriatric PT care and opportunities for MR intervention, this work lays the foundation for future research and systems that can meaningfully improve experiences and outcomes for older adults.

\section{Related Work}

Real-world adoption of technological support for PT remains limited. A survey \cite{merolli_use_2022} revealed that despite willingness to embrace digital technology, particularly given the pandemic \cite{elor_physical_2022, bennell_physiotherapists_2021}, current PT practices primarily rely on face-to-face interactions. While some clinicians use photo capture, patient-logged mobile apps, virtual appointments, and electronic documentation systems, and patients may use wearables like Fitbit or Apple Watch for activity tracking \cite{merolli_use_2022, bailey_digital_2020}, these tools fail to provide the detailed biomechanical patterns observable during in-person assessments. Current apps like Limber \cite{noauthor_limber_nodate} offer remote monitoring and guidance for patients. However, phone apps present accessibility challenges for visually impaired users, and video tutorials lack the personalized guidance of in-person care \cite{kandel_understanding_2024}. MR technology has the potential to deliver individualized guidance to improve patient motivation.

MR systems for motion guidance have been used across various physical domains, including dance \cite{laattala_wave_2024, anderson_youmove_2013}, cycling \cite{kaplan_towards_2018}, basketball free throws \cite{lin_towards_2021}, yoga \cite{jo_flowar_2023}, tai chi \cite{han_my_2017, lee_kinect-based_2014}, martial arts \cite{hoang_onebody_2016}, and physical rehabilitation \cite{doyle_base_2010,garcia_mobile_2014,ayoade_novel_2014, lee_effects_2017}. These systems use spatially-situated visualizations, either directly overlaid on users or mapped to their digital representations, to deliver intuitive and contextually relevant feedback. Comparative studies between MR-based feedback and traditional video demonstrations have consistently shown that immersive feedback  enhances movement accuracy and user engagement \cite{tang_physiohome_2015, anderson_youmove_2013, hoang_onebody_2016}. MR enhances users' perception of depth, height, and size through binocular depth cues and physical movement-based interactions \cite{marriott_immersive_2018, whitlock_graphical_2020, kraus_value_2021, zacks_reading_1998}. While this shows great potential, more research in older patients' and clinicians' needs is required for developing effective MR systems.

MR technology has demonstrated particular promise in medical rehabilitation areas such as stroke recovery \cite{chen_lower_2020, colomer_effect_2016, elor_towards_2018, iruthayarajah_use_2017, jung_envisioning_2022, lohse_virtual_2014, shahid_comprehensive_2023, timmermans_walking_2016} and gait improvements \cite{corbetta_rehabilitation_2015, lee_effects_2017, held_augmented_2020, timmermans_walking_2016}. While one might anticipate older adults struggling to use MR headsets, these applications have shown notable success. For instance, AR-projected obstacles on a treadmill have proven effective for older adult gait and balance practice \cite{timmermans_walking_2016}, while immersive virtual coaching systems have received positive feedback from older adults due to their capability to provide multi-angle viewing perspectives \cite{mostajeran_augmented_2020}. Given these positive indications of usability for older adults, we summarize various applications for MR to assist in geriatric PT, as well as further research avenues for making these systems deployable in daily life. 

%MR telepresence technology enables remote presence and interaction between physically separated individuals \cite{sherman_understanding_2018}. In healthcare contexts, this enables virtual patient-clinician interactions and remote diagnosis \cite{worlikar_mixed_2023}. For example, Tian et al enabled clinicians to remotely examine 3D reconstructive model of the patient in the virtual world while physically feeling the patient's range of mobility using haptic feedback \cite{tian_h-time_2017}. Having access to healthcare from one's home is especially relevant to older adults who may struggle to leave the house. %This included feeling of immersive presence with 3D reconstructions can help make virtual appointments more effective. %In addition to providing benefits for patients, MR could be leveraged for remote clinical assessment and movement analysis. For example, PD-Insigher utilizes AR to allow clinicians to analyze a body and environment reconstruction to understand at home motion behavior. In this paper, we describe different applications of telepresence for both the patient and the clinician, including virtual appointments and clinician motion analysis. 

Prior research has demonstrated MR approaches for motion guidance through various visual representations. Graphical encodings for motion guidance include human models \cite{laattala_wave_2024-1,anderson_youmove_2013,jo_flowar_2023,yang_implementation_2002}, paths and wedges \cite{tang_physiohome_2015,vieira_augmented_2015} and targets \cite{powell_openbutterfly_2020, doyle_base_2010}. Each visual representation of guidance emphasizes different movement aspects: overall body position, motion trajectories, or endpoint goals. In addition to encoding design, some systems use gamification elements to boost engagement \cite{karaosmanoglu_born_2024}. Examples include catching gemstones with butterflies in fantasy environments \cite{elor_project_2019} and participating in simulated sports activities like ball games, shot put, bowling, and balance challenges \cite{reilly_virtual_2021}. However, these design choices present important trade-offs, such as choosing between a fun, gamified, engaging experience, or a more direct approach focusing on precision and user focus \cite{kandel_understanding_2024}. The effectiveness of both visual encodings and gamification elements depends on understanding the users' needs. For older adults specifically, successful design requires deeper insight into their unique motivational challenges, which we identified through our interviews.

\section{Current Challenges}

\begin{figure}
  \includegraphics[width=\columnwidth]{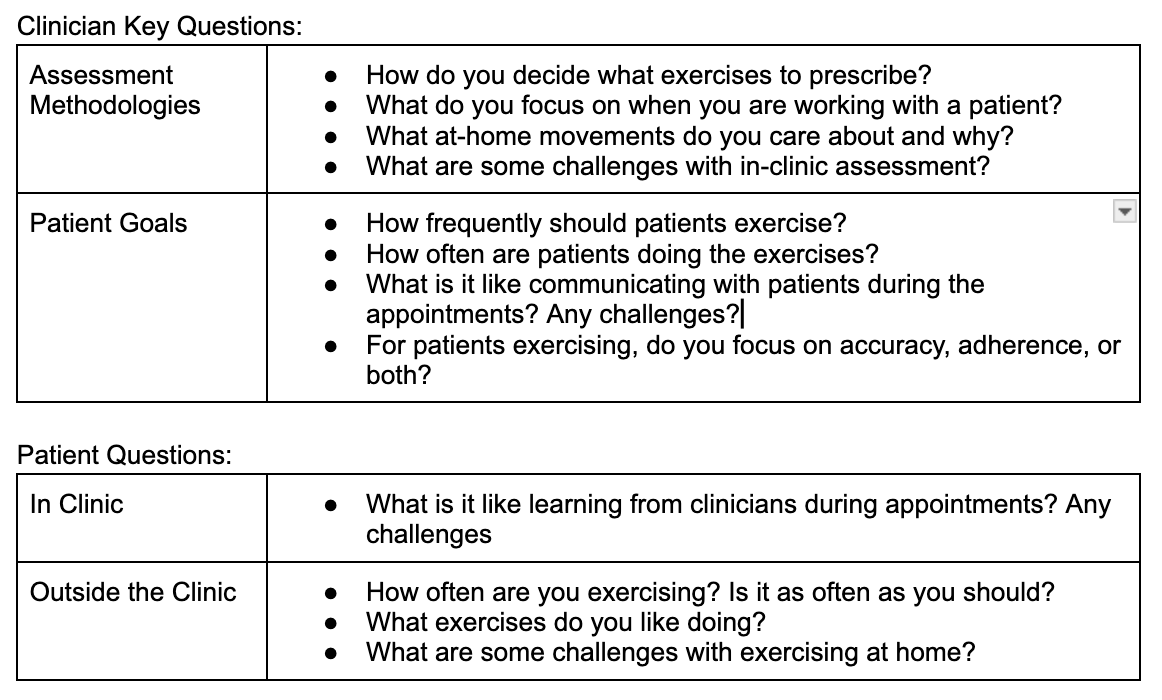}
  \caption{ The interview questions asked to our nine participants, categorized by participant type and question themes.
  }
    \label{fig:questions}
\end{figure}

To investigate the potential role of MR technology in geriatric PT care, we conducted semi-structured interviews with both clinicians and patients. Our participant pool included three licensed PT clinicians (C1-C3; 2 male, 1 female) and six PT patients (P1-P6; 3 male, 3 female). Among the clinicians, two specialized in Parkinson's disease treatment, while the third provided general physical therapy services, primarily to patients over 60 years of age. The patient cohort consisted of five individuals diagnosed with Parkinson's disease and one patient managing chronic back and leg pain. All patients were over 60 years old and engaged in physical therapy treatment at the time of the interviews. While the majority of our participants had Parkinson's disease, participants reported many of their challenges stemmed from the natural aging process rather than their specific condition. We focused our questions for clinicians around their assessment methodologies and patient goals, and our questions for patients around their experiences with PT both in and out of the clinic (Figure 2). 

\subsection{In-Clinic Challenges}

\textbf{Monitoring Patients:} 
Chronic conditions in older adults present unique challenges, as pain, falls, and difficulty performing daily actions can happen at any time in daily life. While effective PT customization requires understanding patients' motor activities and struggles at home, clinicians currently rely only on brief in-clinic observations and patient self-reporting. As one clinician noted, "We rely on patients' recall, which is not always reliable, especially for older patients who are forgetful, or feel embarrassed and lie" (C2). %Given this limited recall, 
Clinicians struggle to know of potential triggers that may explain described motor behaviors of concern. "If a patient told me they fell, that could have been triggered by walking through a door frame, change in carpeting, making a turn, or feet shuffling. I would want to know these things so I can suggest changes, like having support for walking like a walker, or different carpeting " (C1).

\textbf{Analyzing Patients:} 
Clinicians also want insight into complex biomechanical relationships such as precise joint angles and body segment interactions. As one clinician described, "(When evaluating a patient,) I'm thinking about the joint angle: one limb relative to the other... For sit to stands, I'm watching the position of the pelvis, particularly on the way down. When descending we want to make sure it is controlled, velocity of descent is important" (C1). Given the three-dimensional nature of the physical therapy analysis, clinicians can find video replays of movements limiting: "Video's have fixed perspective. If the patient is far away from the camera, I can't really see what's going on. I also may want to look at the other side of the body, but can't" (C2). Limited understanding of potential environment triggers and biomechanical patterns at home impacts the personalized care they can provide for their patient.

\textbf{Teaching Patients:}
During clinical observations, clinicians heavily rely on physical demonstrations, gestures, and touching the patient on relevant body parts to explain biomechanical concepts to patients. Despite advanced age, some patients maintain a strong desire to understand their body mechanics. As P1 expressed: "I like to learn and I have a lot of questions... [the clinician] never explains with visuals, just words." While verbal descriptions provide some clarity, C1 and C2 believed visual feedback was important and could be improved in the clinical session. Older patients may particularly benefit from visual feedback since they face challenges with auditory processing and cognitive speed \cite{lopez-otin_hallmarks_2013}. C2 explained how they try to teach more visually: "Sometimes we show (patients) videos so they can see what they look like. They don't always know what their bodies are doing, like how much they are leaning to the left while walking. Videos work okay, but you can't see the whole body." Patients' limited ability to observe their own movements from multiple angles may impact their capacity for learning.

\subsection{Outside Clinic Challenges}

\textbf{Traveling:} 
Participants P3, P5, and P6 regularly engaged in group physical activities including pickleball, ping pong, aerobics, and boxing classes, emphasizing the social value of group sports. As P3 noted: "I can exercise at home with videos, but I would rather do it with my friend and do pickle ball or boxing." However, transportation and mobility emerged as significant barriers to participation. Fatigue from travel deterred regular attendance, with P6 stating: "Even though it's good for me, it takes a lot of time and energy to get to and from the boxing class, it's very tiring ... I don't go as much as I should." Physical limitations (P2: "It hurts to walk, so I don't leave the house often") and transportation dependency (P4: "a logistical hassle") further constrained participation, highlighting the disconnect between the recognized benefits of social exercise and its practical accessibility for older adults.

\textbf{Remembering Exercises:}
Age-related memory decline impacted older adults' engagement with PT. P1 articulated this challenge: "It's difficult to remember how to do the exercises after the [appointment]. I am getting forgetful. I also forget the multiple exercises... and the number of reps I am supposed to do." Given their forgetfulness, P1 decided to "find some random videos on youtube" to assist with exercises. However, this method for getting one's exercise regimens would not necessarily be personalized to P1's needs. C2 emphasized: "Each patient is different. There is no one regimen that works for everyone. It depends on what that patient needs help with." While digital solutions exist to support exercise adherence and provide personalized instruction, technology preferences among older adults present additional barriers. C3 noted: "Right now we have an app that is great for showing exercises, but it's on the phone. My older patients tell me they don't like working with phones, and will only use it if it's on their computer." This highlights the importance of not only addressing memory support needs but also ensuring technological interventions align with older adults' preferences and comfort levels.

\textbf{Motivation and Adherence:}
Progressive muscle weakness and degenerative conditions like Parkinson's disease create psychological challenges for older adults in maintaining exercise adherence. The emotional toll of aging is evident in P3's reflection: "It's very hard knowing you won't really see progress because [Parkinson's Disease] is degenerative." This lack of visible improvement, combined with age-related fatigue, can impact exercise motivation. P2 described these challenges: "I am not doing [physical therapy] as much as I should. I feel tired, I am not seeing progress, and I get scared that I can hurt myself more." The challenge of diminishing physical capability was particularly evident in P1's description of a "doom spiral," acknowledging that while reduced exercise leads to muscle weakness, the experience of weakness itself made them not want to exercise. C3 described disappointment in the amount of patients who actually do exercises at home, "Only 30 percent of my patients actually do the exercises as much as I want them to." Recognizing these adherence challenges, clinicians adopt a pragmatic approach that prioritizes sustained engagement over technical precision: "I care more about [my patients] being active regularly than the actual activity" (C1). Regular physical activity is crucial in slowing disease progression and preserving muscle strength, yet patients often struggle to stay motivated and exercise consistently despite understanding its importance.
%- adherence
%- motivation

\section{Mixed Reality Opportunities}
 Through synthesis of interview feedback from healthcare providers and elderly patients, we identified several key opportunities where mixed reality (MR) visualization technology could enhance older patient care and outcomes. Given the three-dimensional nature of physical therapy movement and complex sequences of movements timed together, MR shows great potential in providing visual support for clinical analysis and patient guidance.

\subsection{In-Clinic Opportunities}
\textbf{Patient Monitoring:} Current clinical understanding of chronic conditions of older patients is constrained by brief in-person appointments and reliance on patient's forgetful memory, hindering the clinician's ability to create personalized physical therapy regimes for their patients. Motion capture systems %with immersive 3D reconstructions 
could provide clinicians with tools for effective remote body motion analysis, offering valuable insights into how motor challenges manifest in patients' daily lives. %In order to have accurate 3D patient reconstructions suitable for clinical analysis, we require further development in body motion tracking. Existing motion capture technologies present significant implementation barriers for home deployment. Professional systems such as Vicon offer high accuracy but are cost-prohibitive and require extensive sensor placement. Consumer-grade alternatives including Kinect lack sufficient precision and range for comprehensive tracking. While wearable devices enable basic health monitoring, they cannot capture the detailed body and environmental data needed for physical therapy analysis. 
To minimize visual and physical disruption in patient's daily life, future MR headsets could collect data using downward facing cameras and sensors embedded in eyeglasses and a small number of sensors embedded in daily wearables like watches and shoes \cite{zhang_reconstruction_2023, cha_mobile_2021}. Further motion capture development must prioritize patient comfort and privacy. While technologies like lateral-effect photodiodes (LEPDs) \cite{yang_implementation_2002} offer promising solutions for privacy-preserving data capture, more work is necessary to understand and implement patient-centered privacy approaches. Future research should focus on developing comprehensive privacy-aware methods that balance clinical utility with patient autonomy, including privacy-preserving data pre-processing, data capture in only certain environments, and automated anonymization techniques like face blurring in video captures.
%Privacy preservation and patient comfort must be prioritized in future monitoring systems. Although technologies like lateral-effect photodiodes (LEPDs) (citation [66]) offer more privacy-conscious capture methods, additional research is needed to understand privacy requirements and ensure user agency.
 
\textbf{Patient Body Analysis:} 
 While videos and virtual appointments may increase visibility, clinicians indicated that they require understanding of complex biomechanical movements best analyzed with a 3D perspective to best serve their patients. Desktop-based 3D reconstructions require clinicians to learn complex viewpoint manipulation interfaces, whereas MR offers more intuitive spatial rendering that aligns with traditional clinical practices. Systems like PD-Insighter \cite{kandel_pd-insighter_2024} and H-Time \cite{tian_h-time_2017} have demonstrated the potential of immersive reconstructions for patient body analysis (Figure 1.1). Future studies may look to see if what visual representations of the body may be most effective. For example, realistic avatars could improve visualization of biomechanical movements, modeling what is seen in-clinic, but achieving sufficient accuracy and high-renderings currently remains expensive and challenging for clinical settings. This is especially important for older adults, where stuttering feet or tremors are important to capture and analyze. Developing cost-effective methods for generating accurate, coherent body reconstructions that serve clinical needs will be necessary for real-world application.

\textbf{In-Clinic Feedback:}
\begin{figure}
  \includegraphics[width=\columnwidth]{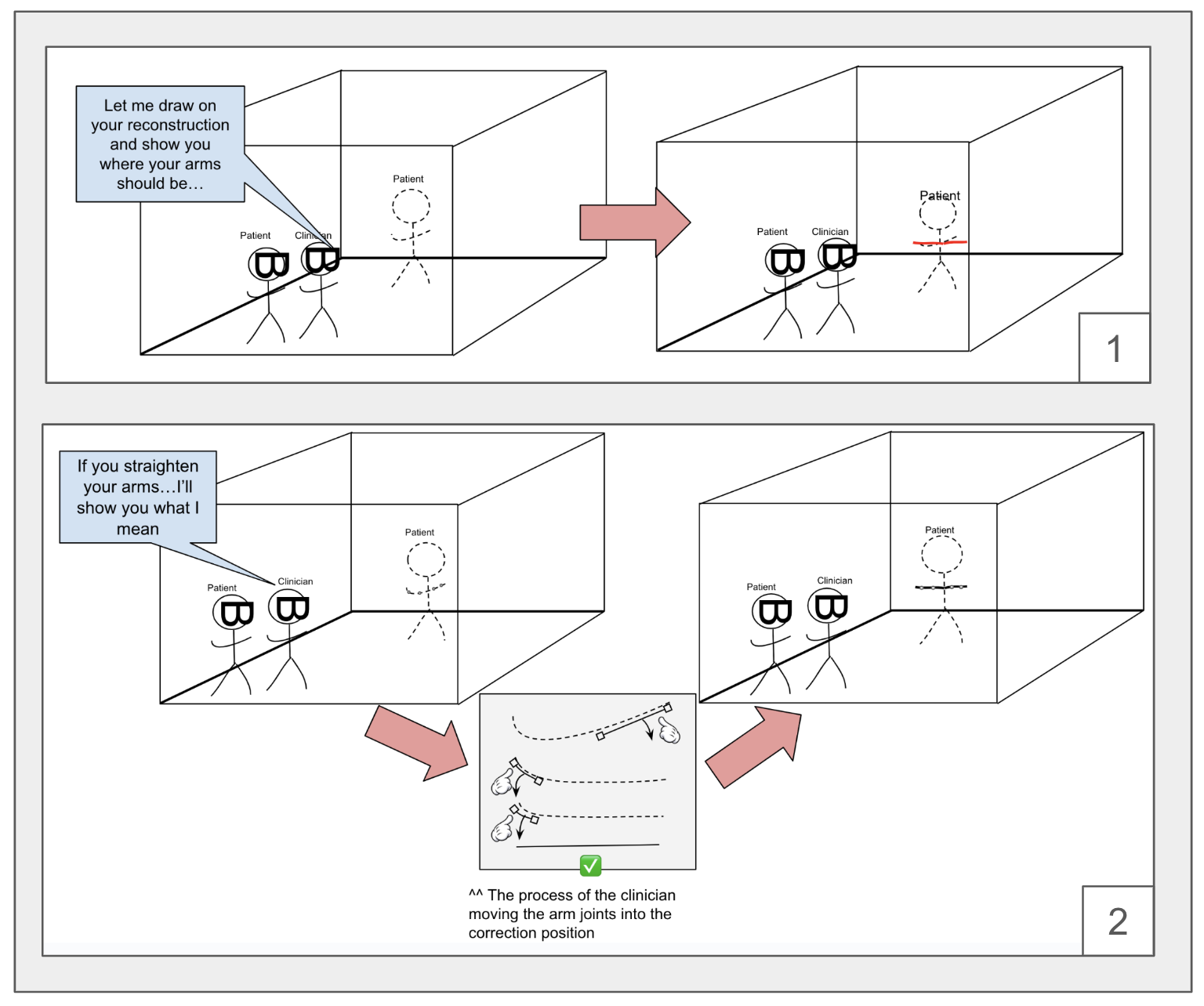}
  \caption{Based on clinicians' suggestions, sketches demonstrating potential methods for clinicians to communicate visually to their patients with shared MR: 1) annotating directly on the reconstruction of the patient, 2) using rigged joints to correct reconstructions body posture.
  }
    \label{fig:questions}
\end{figure}
Given that older patient may struggle more to follow along with verbal feedback \cite{lopez-otin_hallmarks_2013}, clinicians want better visual teaching methods to improve older patient understanding of movement patterns and exercises. Immersive body reconstructions offer opportunities for patients to see their own movements with an analytical perspective (Figure 1.2). Systems could extend beyond simple movement playback to create interactive learning experiences where the clinician describes and interacts in real time with the reconstruction. To help guide older patient's attention and focus, C1 and C2 discussed interactive methods with MR that they felt could be beneficial for teaching older adults about body motion. For example, clinicians could augment 3D reconstructions with real-time annotations, highlighting biomechanical alignments and movement trajectories directly on the patient's virtual body (Figure 3.1). Additionally, clinicians could demonstrate optimal movement patterns by adjusting joint positions and segmental relationships to illustrate corrections (Figure 3.2). Such feedback approaches could enhance patients' understanding of proper movement mechanics and therapeutic goals. Future empirical research should evaluate the efficacy of immersive replays and interactive feedback, with particular attention to the cognitive load and learning outcomes associated with various feedback modalities.

\subsection{Outside Clinic Opportunities}

\textbf{Telepresence for Healthcare:}
Given older adult's mobility and travel constraints, MR systems offer a promising solution to healthcare accessibility constraints through improved remote care delivery in patients' homes. MR telepresence technology enables remote presence and interaction between physically separated individuals \cite{sherman_understanding_2018}. Through telepresence, spatially-aware interactions between patients and clinicians can support real-time virtual appointments where clinicians can observe, assess, and guide patients virtually \cite{tian_h-time_2017}. %The global pandemic highlighted the critical importance of such remote healthcare capabilities, accelerating research into MR telehealth applications (citation Elor).
In addition to visual and auditory communication, tactile feedback is essential for older patients to learn and understand clinician guidance. Haptic feedback integrated in worn sensors could assist in preserving physical touch \cite{worlikar_mixed_2023}, but more research is required for understanding best tactile design practices that optimize for older patient comfort. Beyond individual appointments, telepresence can foster social connection through shared activities \cite{shao_acceptance_2020}. Older patients identified community support as a key factor in exercise motivation, suggesting potential benefits from virtual group sports and game activities. Future research can explore methods for fostering feelings of community and virtual togetherness. Widespread adoption of telepresence requires addressing several challenges, such as privacy protections and data security, mitigating potential psychological impacts of extended virtual immersion, preventing negative social behaviors in virtual spaces, and developing strategies to maintain patient engagement with physical world activities \cite{dwivedi_metaverse_2022}.

\textbf{Motion Guidance:}
\begin{figure}
  \includegraphics[width=\columnwidth]{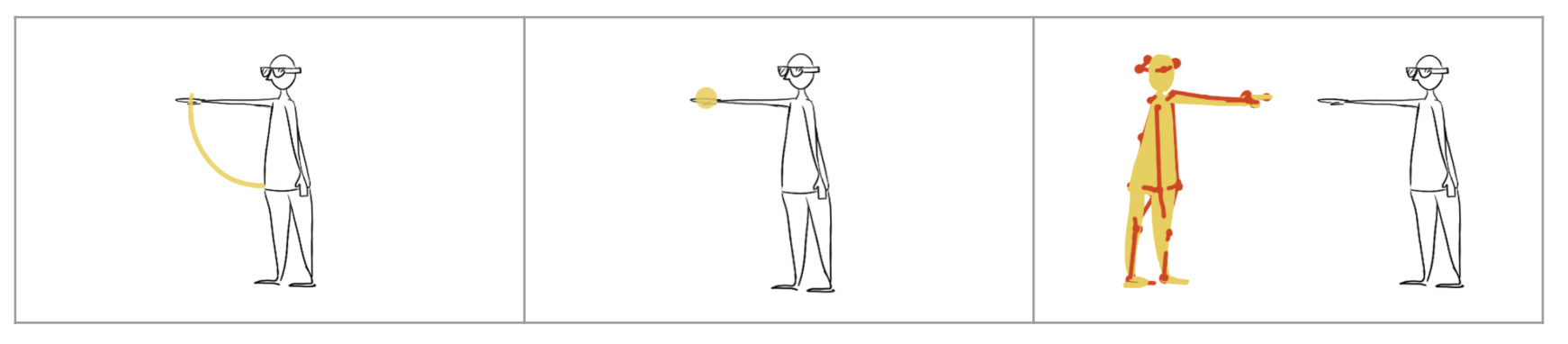}
  \caption{Situating motion guidance with (from left to right) a path, target object, and human model.
  }
    \label{fig:questions}
\end{figure}
Older patients described struggling to remember how to do exercises correctly at home due to memory loss. MR can deliver real-time virtual guidance, situating exercise instruction and feedback with respect to the person's body. Guidance may include following a path, reaching for a target, or copying a demonstrated pose (Figure 4). Older patients also described feeling defeated and unmotivated with their muscles weakening over time, despite exercising. To address this, exercise routines should be tailored to individual patient needs and capabilities. Future systems may explore clinicians leveraging MR platforms to remotely monitor patient progress and dynamically adjust prescribed exercises and repetition counts based on observed performance. This enables clinicians to help target weakened muscles with exercises that are also achievable. Systems could also incorporate adaptive difficulty scaling, automatically adjusting target goals based on real-time biomechanical data to account for experienced muscle fatigue. Dynamic adjustment of exercises and targets helps mitigate patient frustration by ensuring goals remain challenging yet attainable, potentially improving both motivation and adherence. In addition to target placement, designers need to consider the graphical representation of the target. While previous MR systems have emphasized precise movement paths for exercise guidance, our interview findings suggest that for older adults, consistency in daily exercise participation may be more beneficial than perfect movement precision. Easier guidance methods, like reaching for a single target, may be more usable and beneficial for older adults. %Although existing literature provides empirically-validated design guidelines for motion guidance systems, 
Further research is needed to investigate how age-related factors influence the effectiveness of different guidance techniques and how these systems can be optimized for various age demographics. While previous work has demonstrated that headsets were usable for older adults \cite{timmermans_walking_2016,mostajeran_augmented_2020}, a system that older adults are using for long periods of time every day requires hardware development that prioritizes comfort, displays accommodating age-related vision impairments, and control and user interfaces usable for older adults.

\textbf{Gamified Engagement:} While exercise adherence is fundamental for achieving physical therapy outcomes, older adults struggle with exercising frequently despite understanding the benefits. Gamified approaches can help older adults overcome physical and psychological barriers of aging by prioritizing engagement over movement precision, given that regular activity is the primary therapeutic goal for older adults. For example, older adults can slice through floating virtual objects (Figure 1.3), which prioritizes general upper body movement and speed over precise arm location. Game-based interventions can be enhanced through feedback mechanisms that promote feelings of accomplishment and progress. Previous systems \cite{cavalcanti_usability_2019, bell_verification_2019, seyedebrahimi_brain_2019} use text, scores, graphs, or color to communicate feedback. High scores and achievement streaks can serve as motivational tools, encouraging patients to maintain consistent engagement and surpass personal benchmarks. However, the degenerative nature of many age-related conditions necessitates thoughtful feedback design to avoid potential de-motivation. Traditional scoring may reinforce hopelessness when patients struggle to match past performance. Solutions could include periodic score resets and personalized metrics that adjust for condition progression. Future research may explore the emotional impact of feedback and  methods that motivate without patronizing or demoralizing users.

\subsection{Conclusion}
MR can improve both clinician and patient experiences with PT. However, doing so requires understanding of the unique challenges faced by older adults and their clinicians. In this paper, we interviewed three clinicians that primarily work with older adults, and six older adult patients. From their described challenges, we present various opportunities and design considerations for MR to assist in PT patient care. %including patient monitoring and analysis, telepresence for appointments and community, and PT motion guidance and motivation. 
Through highlighting opportunities and further work needed to make MR a real-world solution for geriatric PT challenges and needs, we work towards improving healthcare outcomes for older adults. 

---------
\bibliographystyle{ACM-Reference-Format}
\bibliography{citations}

\end{document}